\documentclass[%
 aip,
 amsmath,amssymb,
 reprint,%
]{revtex4-1}
\usepackage[dvipsnames]{xcolor}
\usepackage{graphicx}
\usepackage{dcolumn}
\usepackage{bm}

\usepackage[utf8]{inputenc}
\usepackage[T1]{fontenc}
\usepackage{mathptmx}
\usepackage{etoolbox}
\usepackage[english]{babel}

\makeatletter
\def\@email#1#2{%
 \endgroup
 \patchcmd{\titleblock@produce}
  {\frontmatter@RRAPformat}
  {\frontmatter@RRAPformat{\produce@RRAP{*#1\href{mailto:#2}{#2}}}\frontmatter@RRAPformat}
  {}{}
}%
\makeatother
\begin{document}

\preprint{AIP/123-QED}

\title[Low Level RF and Timing System Design for the Cool Copper Collider]{Low Level RF and Timing System Design for the Cool Copper Collider}
\author{C. Liu}

 \email{chaoliu@slac.stanford.edu}

\author{A. Dhar}%
\author{M. Breidenbach}%
\author{E. Nanni}%
 \affiliation{ 
SLAC National Accelerator Laboratory, Menlo Park, California, USA.
}%

\date{May 29, 2026.  The following manuscript has been submitted to Review of Scientific Instruments by the AIP. }

\begin{abstract}
The Cool Copper Collider (C\(^3\)) is a linear accelerator (LINAC) concept based on compact, high gradient, and normal conducting accelerator technology to support Higgs boson studies at 250 GeV and 550 GeV center of mass. The C\(^3\) accelerator is ten kilometers in scale and consists of 2,200 RF stations for 550 GeV center of mass. To maintain the stringent beam quality required by the collider across the LINACs, each of the cavities has a dedicated low-level RF (LLRF) system to stabilize the phase and amplitude of the field in the cavities from pulse to pulse and to compensate the  fluctuation of the RF field within each pulse introduced by the beam loading process.  To meet the design goals of being compact and affordable for future accelerators, we have designed the next generation LLRF (NG-LLRF) with a higher integration level based on radio frequency system-on-chip (RFSoC) technology. The NG-LLRF system samples RF signals directly and performs RF mixing digitally. The NG-LLRF has been characterized in loopback mode to evaluate the performance of the system and has also been tested with a standing-wave accelerating structure, a prototype structure for C\(^3\) with peak RF power level up to 16.45 MW. This paper will focus on introducing the LLRF system design and timing system for C\(^3\) and the current NG-LLRF design. The high-power test results at different stages of the test setup with several pulse modulation schemes, including square pulse, pulse with phase reversals, and pulse trains, will be summarized, analyzed, and discussed.
\end{abstract}

\maketitle

\section{Introduction}

The Cool Copper Collider (C\(^3\)) has been proposed as a compact and affordable system for a lepton collider Higgs factory \cite{c3_2025esppu}. We have designed and implemented the low-level RF (LLRF) based on the next generation LLRF (NG-LLRF) system recently developed at SLAC for C\(^3\). The NG-LLRF system is implemented based on the AMD Xilinx radio frequency system-on-chip (RFSoC) technology. The RFSoC family devices integrate high-speed data converters, field programmable logic (FPGA) and processors in a single chip. The integrated Digital-to-Analog Converters (DAC) can directly synthesize RF signal with RF frequency up to 7 GHz, and the integrated Analog-to-Digital Converters (ADC) can sample RF signals with RF frequency up to 6 GHz. The C\(^3\) accelerating structure operates at an RF frequency of 5.712 GHz, which is in the range of direct synthesizing and sampling of integrated date converters in third generation (GEN3) RFSoC devices. 

We developed a range of control and readout systems based on RFSoC devices for radio astronomy instruments operating in the GHz frequency range and evaluated the continuous wave (CW) RF performance \cite{liu2021characterizing, liu2022development,  liu2023evaluating, henderson2022advanced, liu2023higher,liu2024development}. The astrophysics readout typically has extremely stringent requirements in RF performance, and the interleaved structure of the integrated data converters of the RFSoC, which could generate significant interleave spurs, became the primary concern for these applications. However, the ADC demonstrated a spurious free dynamic range over 70 dB and no significant interleave spurs exposed with integrated spectra over 100 s as summarized in \cite{liu2021characterizing} with a first generation (GEN1) RFSoC. The test results in \cite{liu2021characterizing} addressed concerns about RF performance, and then the RFSoC series became a major CW RF platform for a range of physics experiments.  

Normal conducting RF (NCRF) accelerators are typically operated with pulsed RF instead of CW to achieve higher energy efficiency. With the RFSoC based NG-LLRF, the pulsed RF performance was evaluated in both S-band \cite{liu:ipac2025-thps141, liu:icalepcs2025-modr005, liu2026high} and C-band \cite{liu:ipac2024-mocn2, liu2024next, liu:ipac2025-thps140, liu2025high, liu2025compact, liu2025higha}. In \cite{liu2025compact}, the compact LLRF system designed and developed for a compact C-band Linear accelerator (LINAC) as a part of the Advanced Concept Compact Electron Linear Accelerator (ACCEL) program of Defense Advanced Research Projects Agency (DARPA) was described. The ACCEL system has extremely stringent requirements in size, weight, and power consumption (SWaP) for the LINAC. The LLRF design for ACCEL has a high similarity to the LLRF system designed for C\(^3\) that will be introduced in this paper, except that the system for C\(^3\) has a significantly larger scale. The prototype LLRF system developed for C\(^3\) was tested with a high-power stand as described in \cite{liu2025high}. In the high-power test, the pulsed RF that drives the solid-state amplifier and klystron was directly synthesized with the integrated DAC in RFSoC and the RF signals from the couplers at each stage of the test setup were directly sampled by the integrated ADCs in RFSoC. Direct RF sampling or higher Nyquist zone sampling demonstrated pulse-to-pulse amplitude and phase fluctuation levels at approximately 70 femtoseconds (fs) and 0.13\% within 60 consecutive pulses in one second with a loopback setup. The SSA added 80 fs phase jitter, which dominated the overall phase jitter around 150 fs. The phase jitter added by the klystron or RF transmission is under 10 fs. These measurements were taken with an open-loop setup, so the 150 fs phase jitter requirement is highly achievable with a closed-loop LLRF system \cite{liu2025high}. In summary, the pulsed RF stability of RFSoC is considerably higher than the requirement of C\(^3\).

The LINAC design of C\(^3\) has evolved and settled in the last several years. Based on the latest LINAC design, the LLRF architecture was designed, including timing and synchronization, for the entire LINAC. The scaling up plan and LLRF system costs will be summarized in Section \ref{system}. The NG-LLRF was adapted to different applications and some of the versions involve redesign of RF data converter configurations and firmware, and optimization of power level in the RF chain. Based on the LLRF system designs for C-band LINACs, we have designed a new NG-LLRF chassis as the latest version of the prototype for C\(^3\), which will be introduced in Section \ref{chasis}. The high-power test results of NG-LLRF with the prototype C\(^3\) were mostly summarized in \cite{liu2025high}. In this paper, the high-power tests performed at a peak power level of 16.45 MHz with different pulse modulation schemes will be summarized and discussed in Section \ref{hp}. The tests include phase flipping driven and measured by NG-LLRF, which is an essential modulation technique used for SLAC Energy Doubler (SLED) \cite{farkas1974sled,lin2022x} type pulse compressors.  

\section{System Architecture Design} \label{system}

\begin{figure*}
\includegraphics[width=\textwidth]{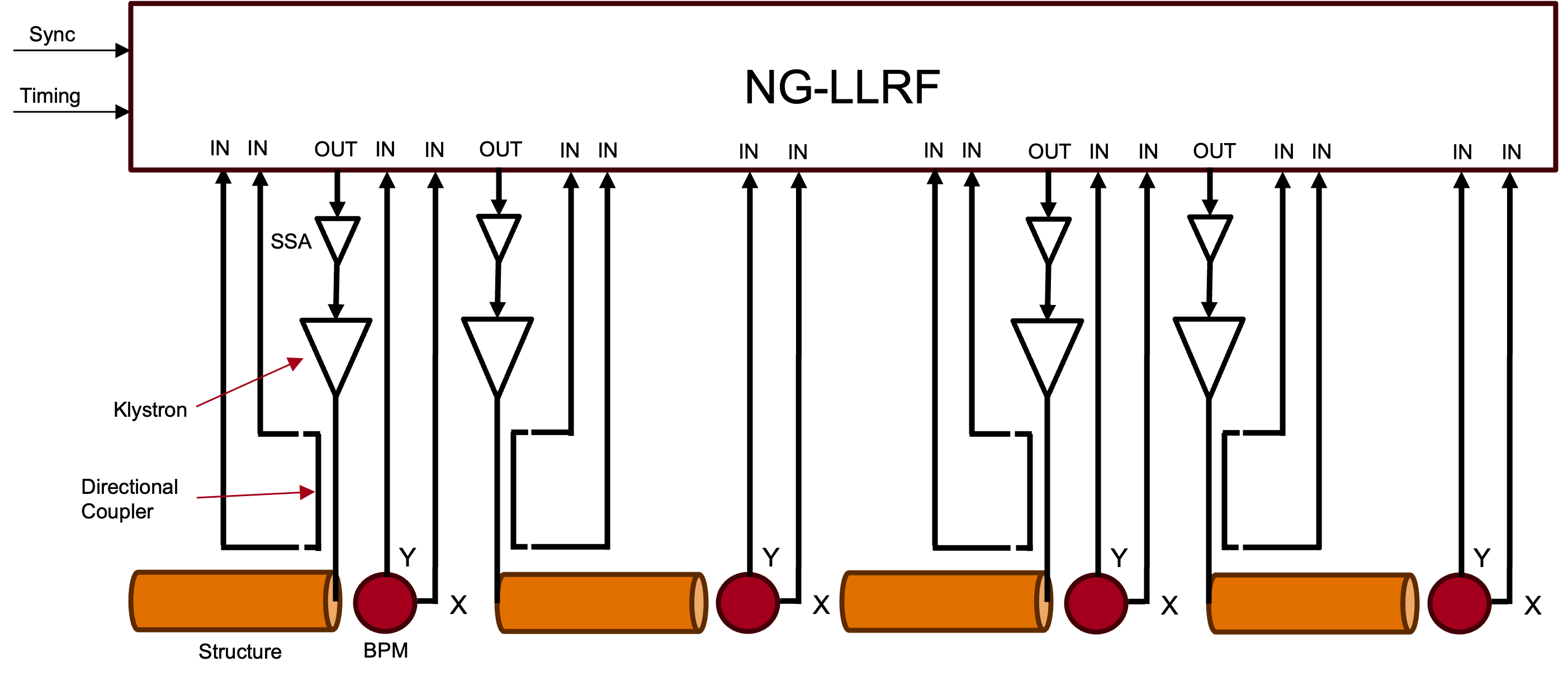}%
\caption{The conceptual block diagram of a single NG-LLRF module for C\(^3\). The NG-LLRF system has 16 RF inputs and 16 outputs. All the RF inputs are utilized and only 4 of the RF outputs are used to drive the 4 accelerators for half of a cryo-module.}\label{fig-1}
\end{figure*}

Figure \ref{fig-1} shows the conceptual design of an NG-LLRF module for C\(^3\) linear accelerators (LINACs).  The NG-LLRF is designed to be synchronized with the RF reference and is triggered by the accelerator timing system. We are working on more detailed designs of the synchronization and timing systems considering scaling up for the entire collider. As C\(^3\) LINAC has a high channel density of RF inputs and outputs, the RFSoC device with 16 ADCs and 16 DACs has been identified as the base platform of NG-LLRF for it. Each NG-LLRF module can drive and monitor 4 accelerators. For each of the accelerator, there are 4 RF signals, which are forward and reflection couplers from an accelerating structure, and the two RF signals from the beam position monitor (BPM) for each structure. Based on each pair of forward and reflection RF pulses directly sampled by the integrated ADCs in RFSoC, an updated RF pulse will be computed in firmware and a DAC will synthesize the new RF pulse directly. In \cite{liu2024next,liu2025compact}, the architecture of feedback loop designs is described in more detail.

In Table \ref{tab-1}, the breakdown of the subsystems at different levels of C\(^3\) is summarized. Based on these statistics, the C\(^3\) LLRF system requires 17,600 RF inputs and 4,400 RF outputs in total. Therefore, around 1,100 NG-LLRF modules will be required for the entire collider. The estimated cost of an NG-LLRF module is currently around \$20k. For each module, we will be using 16 inputs and 4 outputs from each of the NG-LLRF channels. Therefore, the estimated cost per RF channel is approximately \$1k, which is cost-effective compared to the conventional LLRF system. The total cost of the LLRF system will be approximately \$22M excluding the timing and synchronization systems and cable networks.

\begin{table}
\caption{\label{tab:table1}System breakdown of the C\(^3\)}
\begin{ruledtabular}
\begin{tabular}{lcr}\label{tab-1}
System Level &  Number of Modules   \\\hline
LINACs        & 2   \\
Cryomodules   & 275 \\
Rafts         & 4   \\
Accelerators  & 2   \\
RF inputs     & 4   \\
RF outputs    & 1   \\
\end{tabular}
\end{ruledtabular}
\end{table}

\section{NG-LLRF Prototype Chassis} \label{chasis}

\begin{figure}
  \begin{center}
  \includegraphics[width=3.4in]{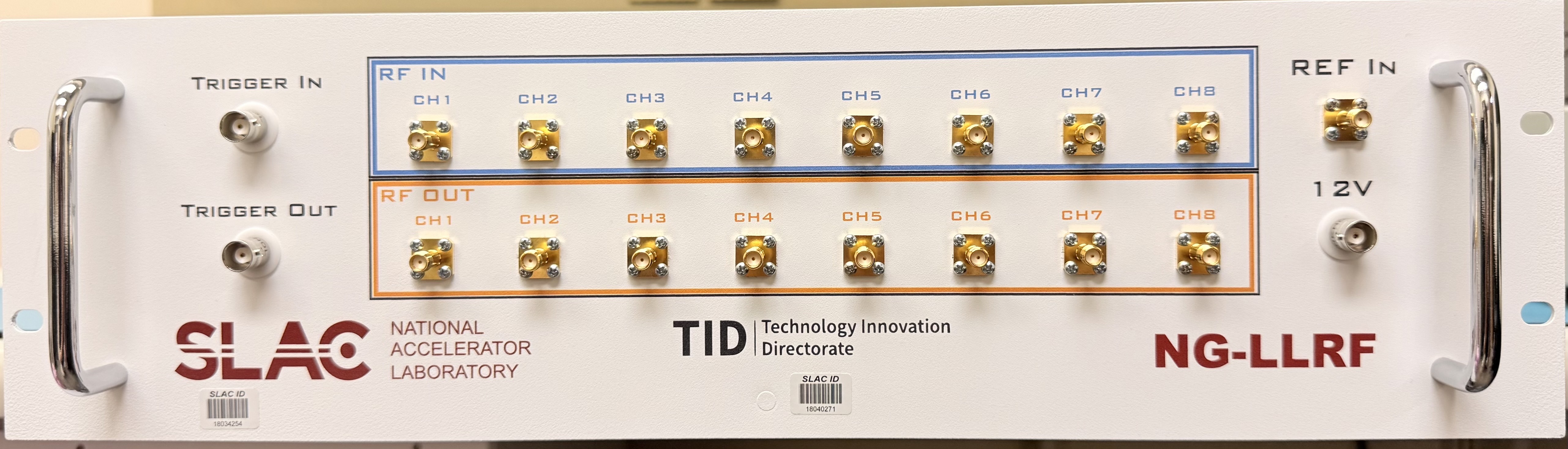}\\
  \caption{The front panel of the prototype NG-LLRF chassis.}\label{fig-2}
  \end{center}
\end{figure}

We designed and fabricated an NG-LLRF chassis as a prototype for C\(^3\). This version of the NG-LLRF chassis can have up to 8 RF input channels and 8 RF output channels, which are accessible from the front panel of the chassis as shown in Figure \ref{fig-2}. The NG-LLRF can be triggered by external trigger source by the "TRIGGER IN" port, and it can also be the trigger source for other devices via the "TRIGGER OUT" port. Multiple NG-LLRF chassis can be synchronized to a common RF reference by injecting the reference to the "REF IN" port. There is a BNC port marked as "12 V", which is reserved for the power supply of the low noise amplifier for the RF drive if it needs a separate power supply with lower noise.   

\begin{figure}
  \begin{center}
  \includegraphics[width=3.4in]{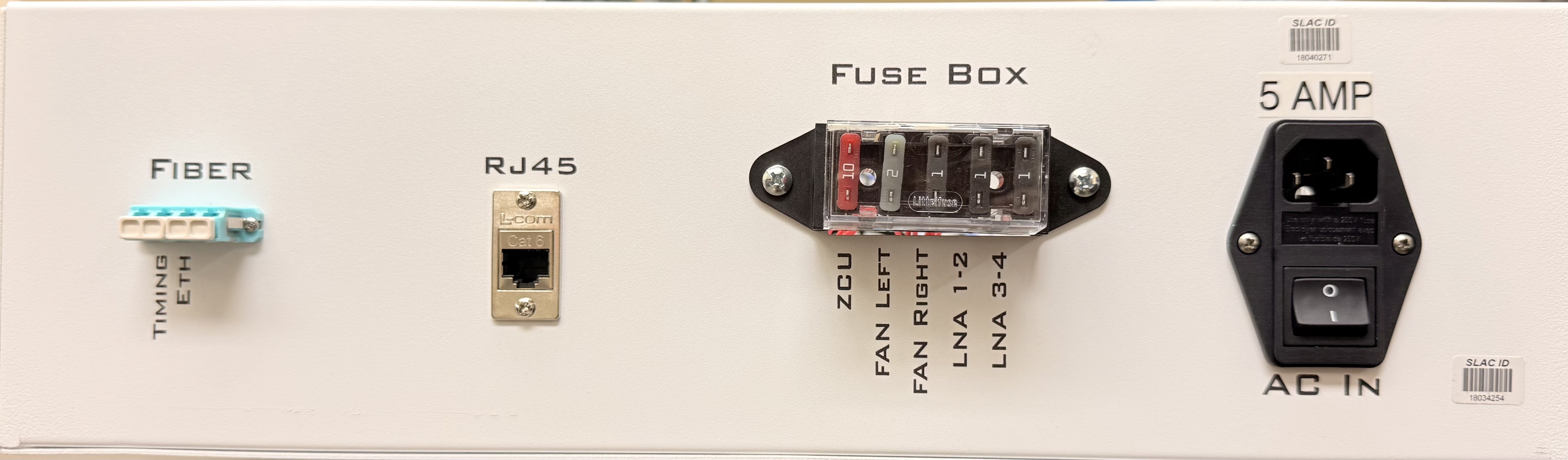}\\
  \caption{The back panel of the prototype NG-LLRF chassis.}\label{fig-3}
  \end{center}
\end{figure}

The communication interfaces are located on the back plane of the chassis, as shown in Figure \ref{fig-3}. The port labeled as "RJ45" is the Gige-bit Ethernet (GbE) connection, for software configuration, low speed data transmission, and remote update of firmware and software images. Two fiber ports are reserved for timing interface and 40 GbE for high speed data transmission. There is a fuse box on the back plane and the components in the chassis are fused separately to offer individual protection. The chassis can be powered directly by a main outlet through the "AC IN" socket, which is also fused. 

This prototype chassis was designed with an RFSoC commercial evaluation board, ZCU208 \cite{zcu208}, which integrates eight 14-bit 5 Giga Samples per Second (GSPS) ADCs, and eight 14-bit 10 GSPS DACs. With the firmware configuration using the internal digital mixers in RFSoC, the maximum sampling rate for the integrated DACs have been reduced to 7 GSPS, and the maximum RF output frequency is 7 GHz. The integrated ADCs have a maximum RF input frequency of 6 GHz. The 5.712 GHz operating frequency of C\(^3\) is within the direct sampling range for both ADCs and DACs. The output of DACs and input of ADCs are routed to a FPGA Mezzanine Card (FMC) connector as differential pairs. An FMC daughter board included in the evaluation kit, that breaks out the differential pairs, is plugged into the ZCU208 main board. The breakout differential pairs are converted to single-ended signals with RF baluns on the daughter cards through breakout cables. The baluns on the daughter card have different frequency ranges, and only four of them cover the C-band frequency. For this prototype chassis, one of the DACs was wired up to the front panel to generate the RF signal that drives a test stand, and three of the ADCs were wired to the front panel to measure the RF signals directly. For the final NG-LLRF chassis, the ADCs and DACs will be wired in a four to one ratio, as mentioned in Section \ref{system}. A custom daughter board with differential pairs directly routed to identical baluns that covers the RF frequency up to 6 GHz was designed and we will use a daughter board based on this design in the final LLRF system of C\(^3\). 

The maximum output power is 1 dBm with the setup, and the balun and cabling have some additional loss. Therefore, a low noise amplifier, ZX60-83LN12+ from Mini-Circuit, which has approximately 20 dB gain in the C-band frequency range, was used to amplify the RF signal generated by the DAC to have more dynamic range to drive the solid-state amplifier at an appropriate power level. The trigger signals are generated or received via the general purpose input and output (GPIO) channels that were broken out on the same daughter board. Depending on the system setup, the NG-LLRF can be the source for triggers of other components and/or triggered by external signals. The trigger signals in both ways were level shifted to 3.3 V by a custom trigger board. 

The clock system of the NG-LLRF can run either stand alone with an on-board oscillator or be synchronized to an external reference via the "REF IN" port on the front panel. By programming the LMK04828B Phase-Locked Loop (PLL), the clock module mounts on the main board can be synchronized with external clock at different frequencies and generate the clock signals at desired frequencies for the FPGA and the integrated data converters. For this version of NG-LLRF, the LMK04828B generates an intermediate reference clock at 245.76 MHz, which is propagated to the data converter tiles in RFSoC. The reference clocks are distributed with a configurable network in RFSoC, and the integrated PLLs in RFSoC convert the reference clocks to high clocks for the data converters.   

The entire chassis is powered by a power supply module mounted on the side wall, which takes the main supply and generates the DC supplies for the components. It generates 12V supplies for the ZCU208 main board and the amplifier, and 5V for fans that dissipate the heat generated by the main board. 

As mentioned in Section \ref{system}, the RFSoC with 16 ADCs and 16 DACs are selected when scaling up the NG-LLRF. This prototype was built with synergy with the S-band LLRF system upgrades the LLRF for the Next Linear Collider Test Accelerator (NLCTA) and test laboratory at the SLAC National Accelerator Laboratory, which have lower channel density. However, most of the hardware designs and all the firmware and software designs are portable between the two RFSoC device families.

\begin{figure}
  \begin{center}
  \includegraphics[width=3.4in]{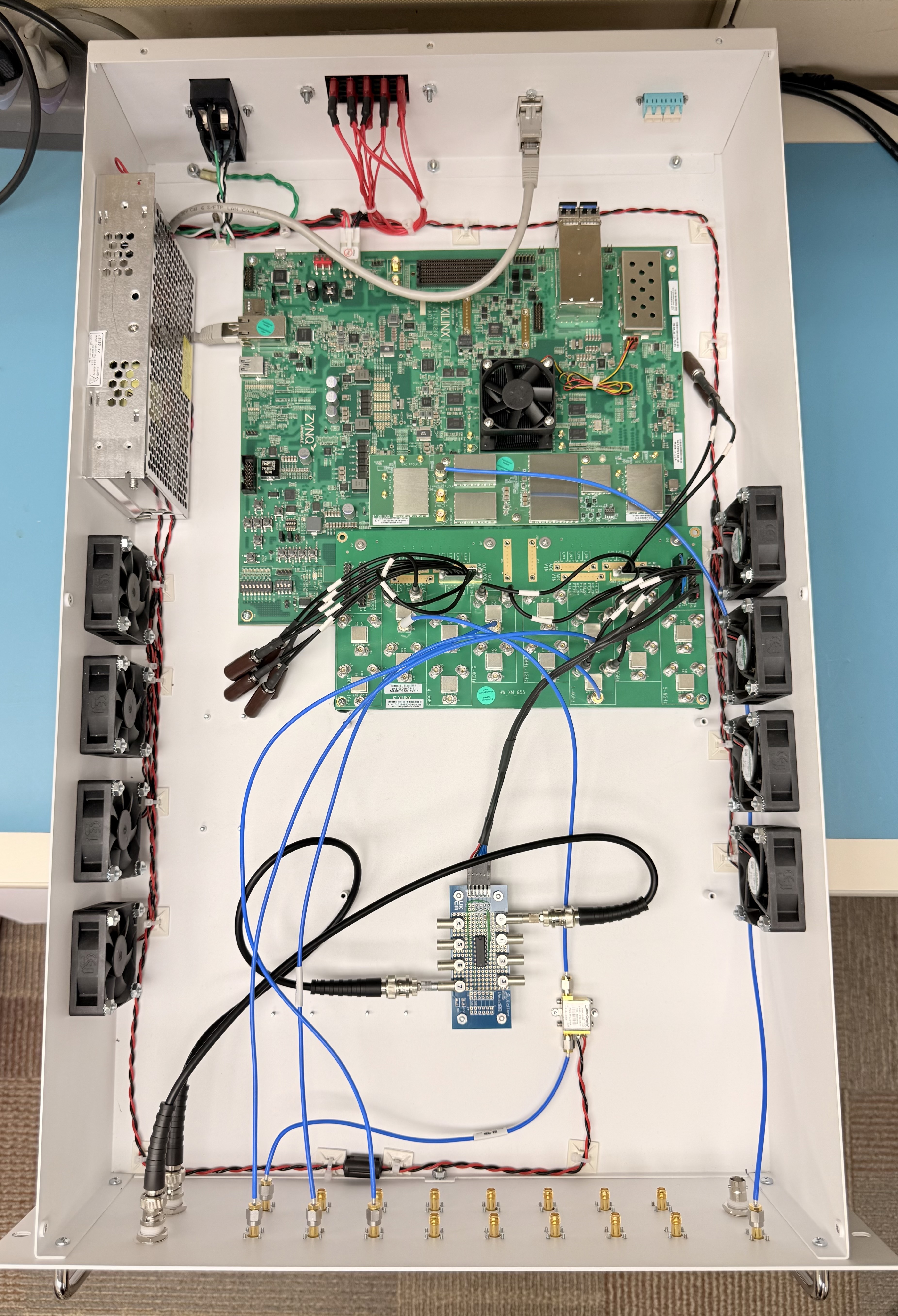}\\
  \caption{The inside layout of the prototype NG-LLRF chassis.}\label{fig-4}
  \end{center}
\end{figure}


\section{High-power Tests} \label{hp}

A prototype NG-LLRF was integrated with a C-band high-power test stand that drives a prototype C\(^3\) structure and the test setup was introduced in \cite{liu2025high}. The RF stability performance and the RF signals of the test setup are also summarized and analyzed in \cite{liu2025high}. The NG-LLRF demonstrated high flexibility in shaping RF pulses and high precision in measuring RF signals from couplers, and a selection of pulse shaping test results was summarized in \cite{liu2025high,liu:ipac2025-thps140}. For future C\(^3\) designs, accelerator control can be transformed from manual intervention to autonomous operation based on artificial intelligence and machine learning (AI/ML) to improve overall reliability and operation efficiency. Therefore, the flexibility of the LLRF in shaping and measuring the RF pulses at high power levels becomes even more critical. In this section, the test results with peal RF power at 16.45 MHz with both magnitude and phase modulation schemes will be discussed.

\subsection{Pulse Train Modulation}

The pulse width is one of the primary parameters that defines the RF characteristic and efficiency of an high power RF system. In this set of experiments, two RF pulse modulation schemes will be compared, one with RF power on for 450 ns, which is the same as the on time of the modulator, and the other with an approximately 250 ns RF pulse. The high power test stand was driven by an RFSoC based LLRF system as introduced in \cite{liu2025high}. The maximum peak power achieved at the stage of high power test stand conditioning in this experiment at the repetition rate of 60 Hz within the desired break down rate was approximately 16.45 MW with the modulator power for approximately 450 ns. Therefore, only one full 250 ns RF pulse has been switched on in each of the RF trains. For RF pulses with lower peak power levels, such as 4.2 MW or 5.7 MW, the modulator can be powered on for approximately 1 \(\mu\)s and at least two 250 ns RF cycles can be completed in each of the RF pulse trains. A selection of test results for lower power has been summarized in \cite{liu2025higha} and the discussion in this section will focus on the test results taken at the maximum RF power.
\begin{figure}
  \begin{center}
  \includegraphics[width=3.4in]{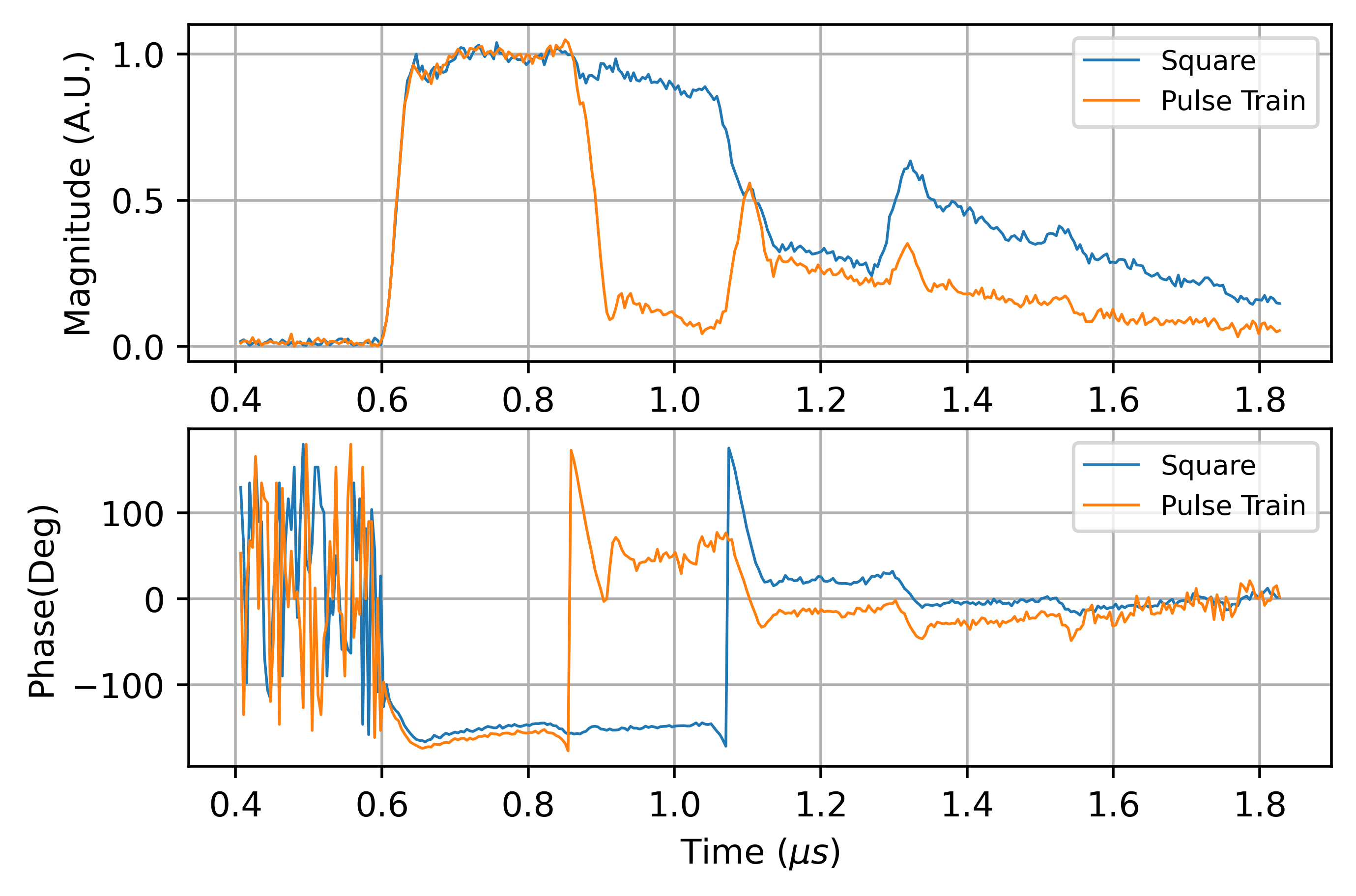}\\
  \caption{The magnitude and phase of baseband pulses from the klystron forward coupler of high power test stand driven by RF pulses modulated with square and square wave with pulse train every 250 ns envelopes. The modulator is powered on for 450 ns and the peak power injected to the prototype structure is around 16.45 MW.}\label{fig-5}
  \end{center}
\end{figure}

Figure \ref{fig-5} compares the forward RF signals of the two modulation schemes. In the ideal case, the magnitude level of pulses from froward coupler for both 450 ns and 250 ns RF pulse widths should drop rapidly after the RF is switched off. However, the forward pulses shown in Figure \ref{fig-5} still have high residuals after the RF is off, which can be mainly attributed to the cross coupling of the reflected power to the forward coupler. Similar cross coupling has also been observed and discussed in \cite{liu2025high}, which can be improved by optimizing the coupler matching design or by using software algorithms to extract the true forward RF signal. For a system driven by RF pulse trains, the interactions between the forward and reflected RF fields become more complicated than square RF pulses. The forward coupler is often the only port for the feedback stability control loop of C\(^3\) structure. Therefore, we should improve the accuracy for measuring the forward pulse shape if we decide to use more exotic RF waveforms to drive the system. 

\begin{figure}
  \begin{center}
  \includegraphics[width=3.4in]{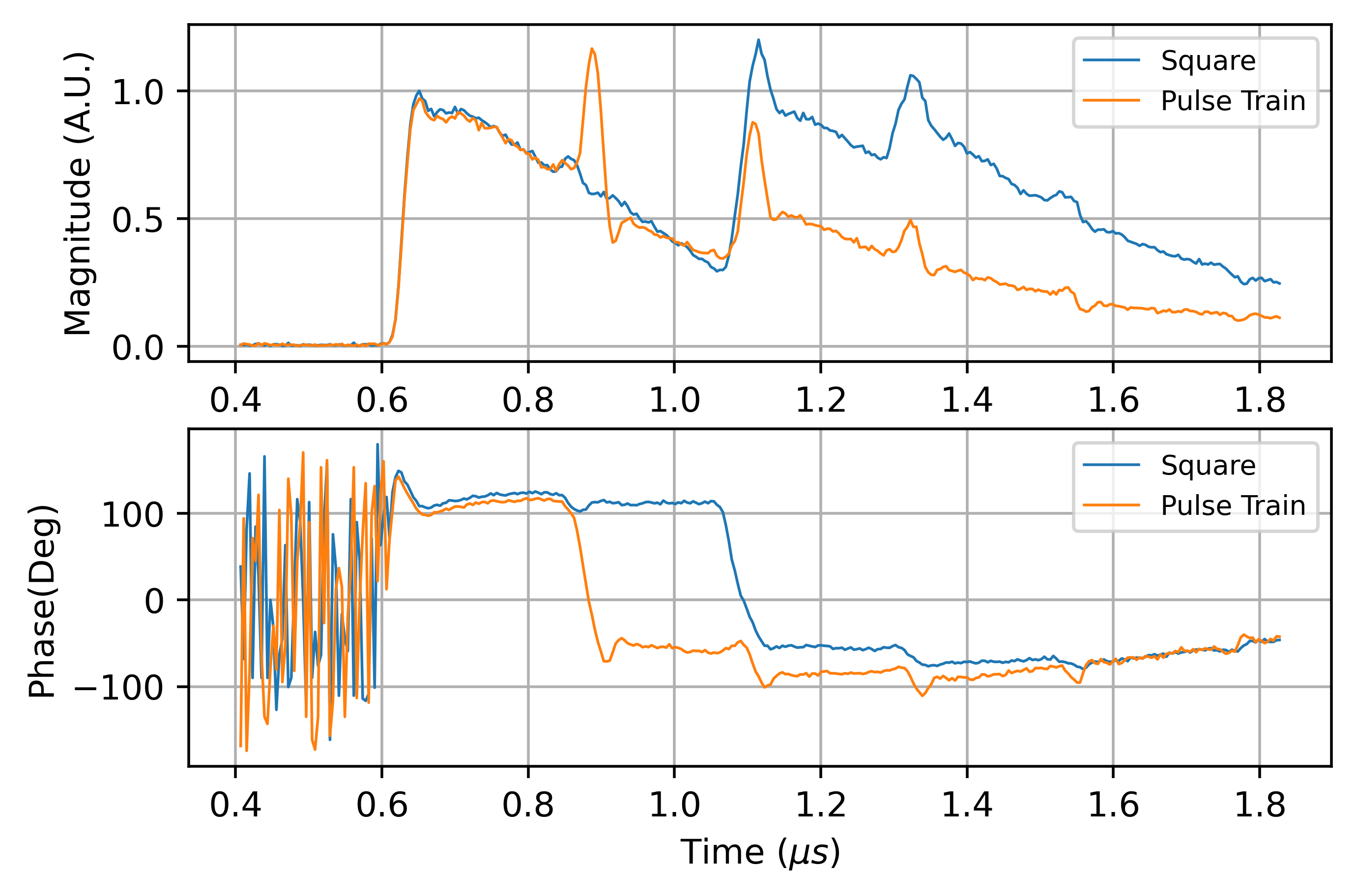}\\
  \caption{The magnitude and phase of baseband pulses from the reflection coupler of the high power test stand driven by RF pulses modulated with square and square wave with pulse train every 250 ns envelopes. The modulator is powered on for 450 ns and the peak power injected to the prototype structure is around 16.45 MW.}\label{fig-6}
  \end{center}
\end{figure}

Figure \ref{fig-6} compares the reflections signals of the two modulation schemes. The field filling magnitude and phase traces for the two modulation schemes are highly correlated for the initial 250 ns after the RF is switched on. At the moment the RF is turned off for the 250 ns pulse train, the phase of the reflection signal is reversed in approximately 40 ns, which indicates the power flows in reversed direction since the RF is off. The magnitude of the reflection signal spikes in about 20 ns and then descends as the RF power dissipates in the waveguide. When the modulator is off, there is another peak in magnitude of the reflection signal, which is the characteristic of the critical coupled design of the prototype C\(^3\) structure used in this case. As the RF power dissipates in the waveguide, it reflects back and forth between the klystron and the structures several more times, which is picked up by the reflection coupler and appears as the following peaks on the reflection signal magnitude traces. The positions of the peaks in the reflection traces are highly correlated with the peaks in the forward traces. The cross coupling can be subtracted by applying the methods used for the square pulses in \cite{liu2025high}.

\subsection{Phase Reversal Modulation}

Phase modulation is another typical technique used for RF stations across accelerators. In \cite{liu2025high}, we discuss the test results with a linear phase ramp modulation, which is equivalent to drive the structure off resonance. The NG-LLRF and the klystron successfully generated the high power pulses with the modulation scheme, but the majority of RF power was reflected back. RF pulse modulated with a phase reversal or flip of 180 degrees is commonly used since SLAC Energy Doubler (SLED) was designed and implemented \cite{farkas1974sled,lin2022x}. 

\begin{figure}
  \begin{center}
  \includegraphics[width=3.4in]{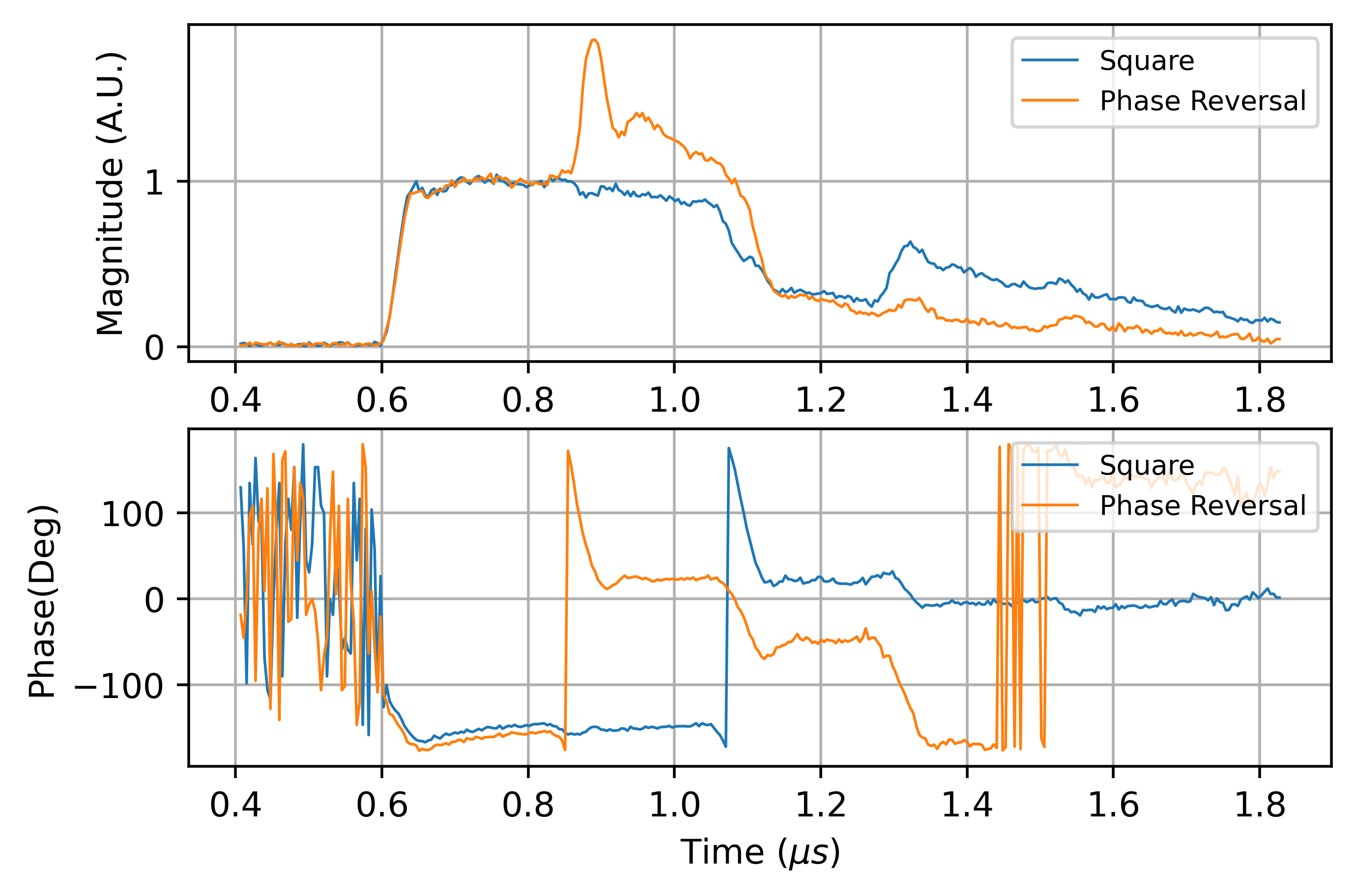}\\
  \caption{The magnitude and phase of baseband pulses from the forward coupler of the high power test stand driven by RF pulses modulated with a square pulse and a square pulse with a phase flip every 200 ns. The modulator is powered on for 450 ns and the peak power injected to the prototype structure is around 16.45 MW.}\label{fig-7}
  \end{center}
\end{figure}

\begin{figure}
  \begin{center}
  \includegraphics[width=3.4in]{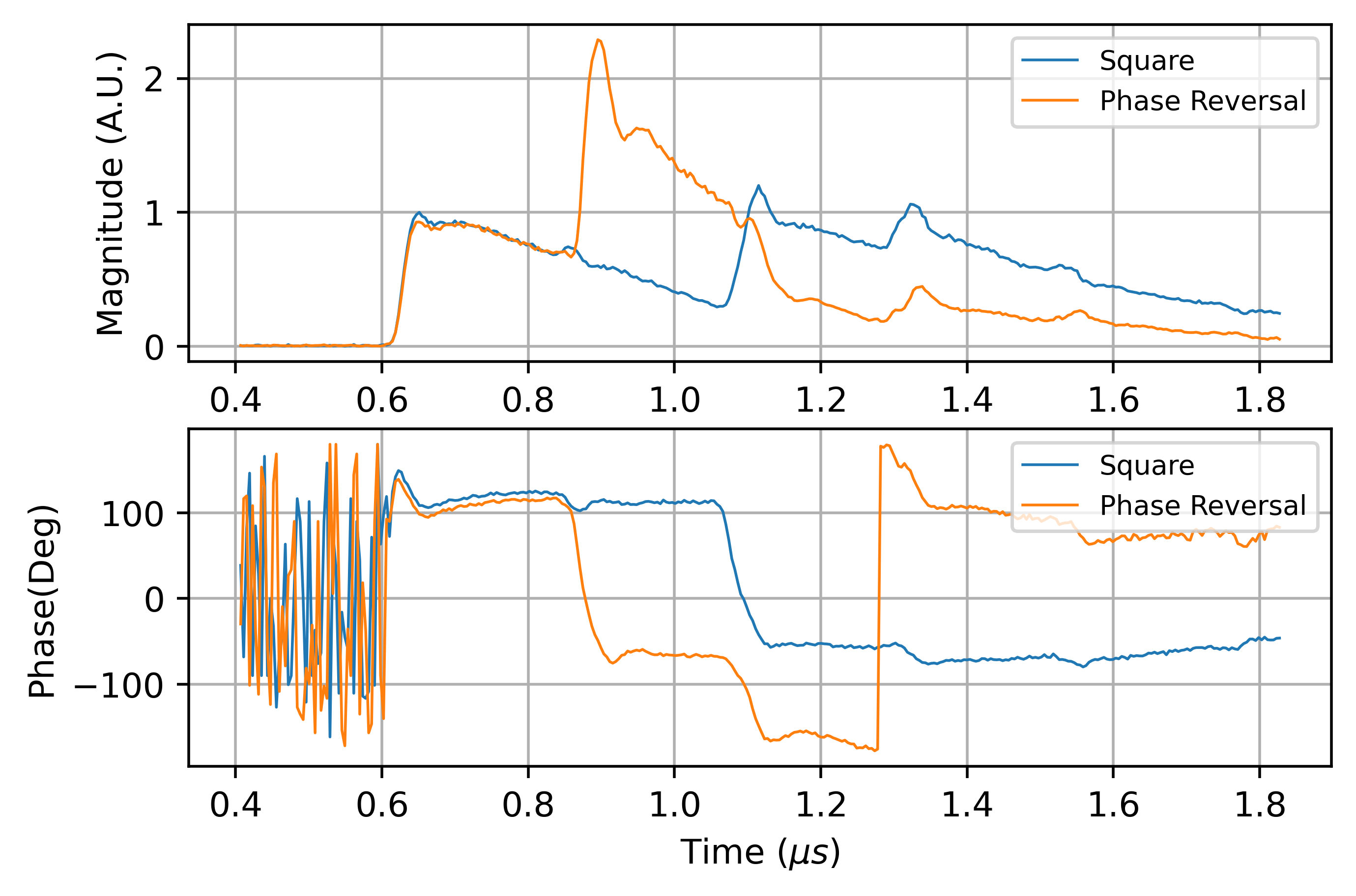}\\
  \caption{The magnitude and phase of baseband pulses from the reflection coupler of the high power test stand driven by RF pulses modulated with a square pulse and a square pulse with a phase flip every 200 ns. The modulator is powered on for 450 ns and the peak power injected to the prototype structure is around 16.45 MW.}\label{fig-8}
  \end{center}
\end{figure}

In this section, the baseband pulses measured with the test stands driven by RF pulses modulated with a square pulse and pulse with a phase flip will be compared. In this test, the peak power of the RF pulse injected into the structure is approximately 16.45 MW. The duration of the RF pulses is approximately 450 ns and the phase flip occurs every 200 ns. As Figure \ref{fig-7} shows, the forward signals of the two modulation schemes are highly matched in the 200 ns after the RF is turned on, but there is a dramatic increase in the magnitude when the phase of the RF drive is flipped.  For the RF drive with phase flip, the magnitude remains constant throughout the pulse. And the profile of the forward signal after phase flip is highly correlated with the reflection signal shown in Figure \ref{fig-8}. Therefore, the spike in the forward signal can be primarily attributed to the cross-coupling from the reflection coupler to the forward coupler, which is similar to what we observed for square RF pulse drive in \cite{liu2025high}. 

As Figure \ref{fig-8} shows, the magnitudes of the reflected signals are scaled to the first peak when driven by RF pulses with square wave modulation. When the phase is flipped, the test setup demonstrates a pulse compression rate of 2.2 even though the structure is not designed as a pulse compressor. 

\begin{figure}
  \begin{center}
  \includegraphics[width=3.4in]{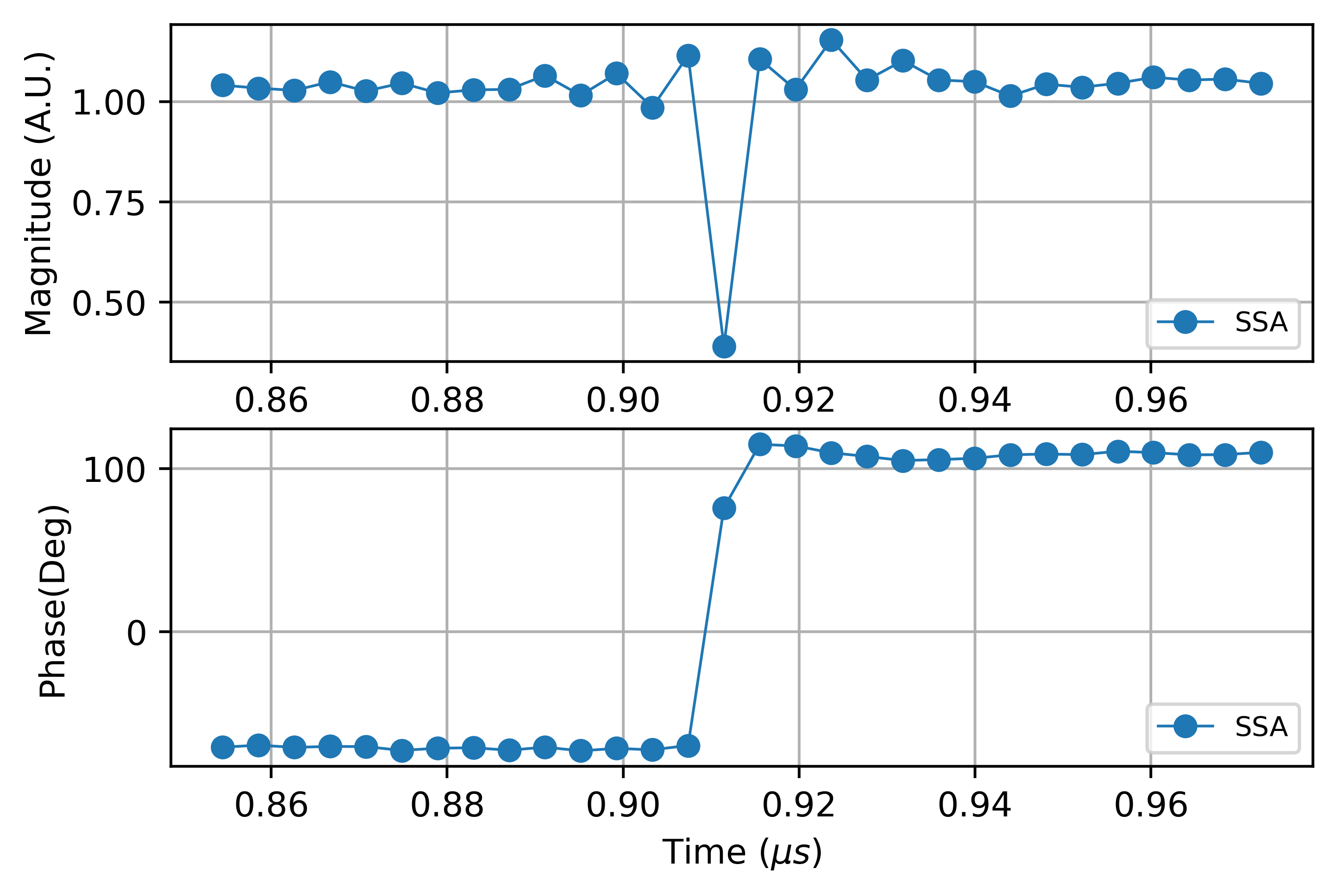}\\
  \caption{The magnitude and phase of the SSA output in 100 ns around the phase flip.}\label{fig-9}
  \end{center}
\end{figure}

\begin{figure}
  \begin{center}
  \includegraphics[width=3.4in]{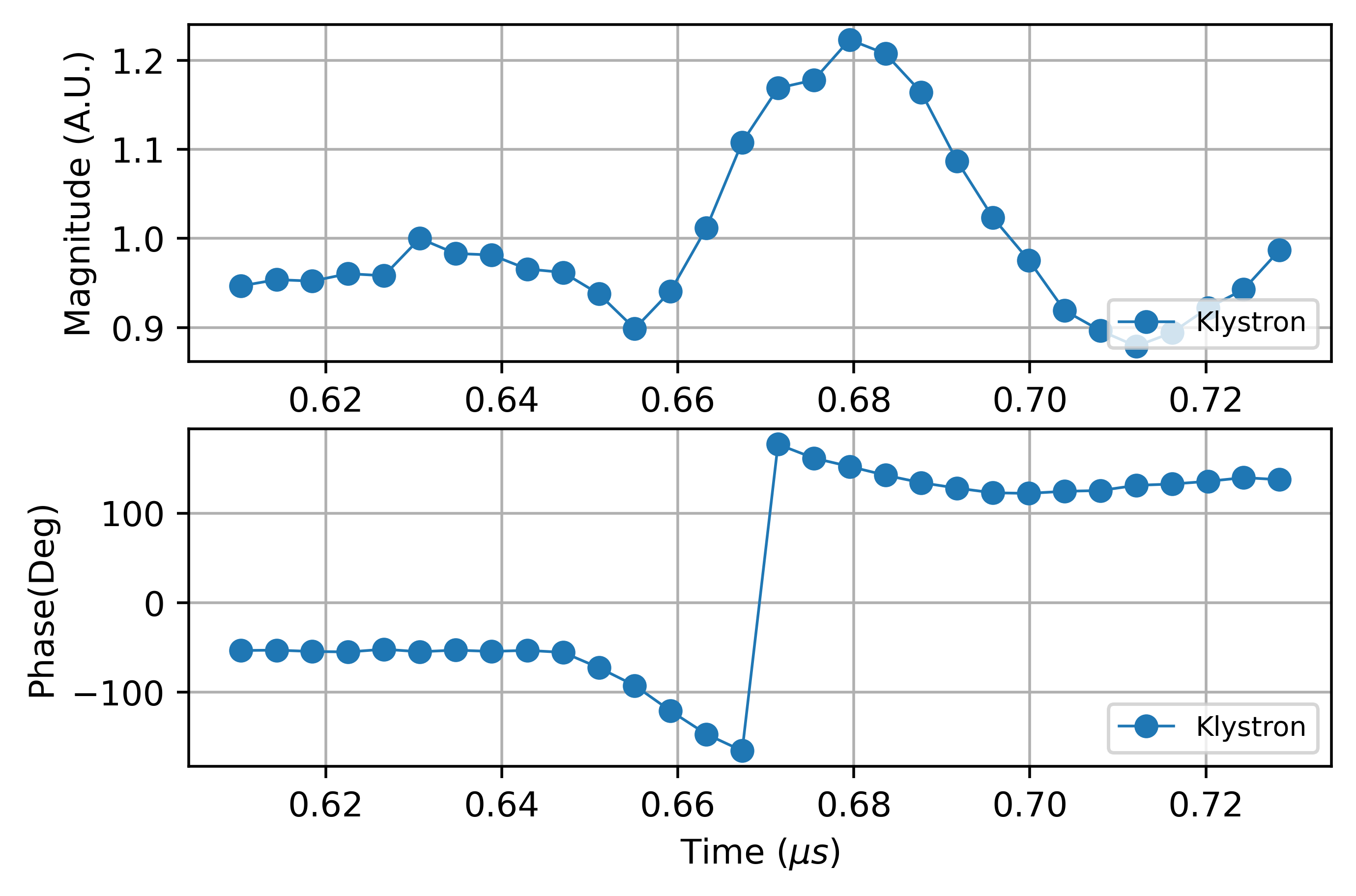}\\
  \caption{The magnitude and phase of the klystron forward in 100 ns around the phase flip.}\label{fig-10}
  \end{center}
\end{figure}

For a pulse compressor, the flip time of the phase has a significant impact on the compression rate. We measured the flip time at the multiple stages of the high power test stand. Figure \ref{fig-9} and \ref{fig-10} show the magnitude and phase in approximately 100 ns time intervals centered on the phase flips at the output of SSA and the klystron forward coupler, respectively. In this test, the peak output power level of the SSA is approximately 1 kW and the phase was flipped in approximately 8 ns as shown in Figure \ref{fig-9}. The phase flip introduced minor fluctuations on both magnitude and phase and settled in 10 ns. However, phase flip introduced more significant fluctuations in the magnitude and phase of the klystron forward signal, as shown in Figure \ref{fig-10}. The phase declines for 20 ns and settles in 20 ns after the phase flips, which is due to the narrow bandwidth of the klystron. The phase ramp on the pulse tail of the klystron forward can reduce the peak power as a pulse compressor. For an RFSoC based LLRF system, the RF pulse modulation is fully implemented in digital domain, so the RF pulse can be shaped with high flexibility and precision. We can explore and experiment with different phase and magnitude modulation schemes to optimize the peak power of the pulse compressor or for other design goals.

\section{Conclusion}
 
The higher integration level and wide RF bandwidth of the data converters of RFSoC revolutionized the architecture of LLRF control system for particle accelerators. Based on the latest conceptual design, the C\(^3\) RF structures will need at least 7,600 RF inputs and 4,400 RF outputs, which can be realized with just 1,100 NG-LLRF modules we describe in this paper. Compared with conventional LLRF control systems for particle accelerators, the NG-LLRF has significant advantages in SWaP and cost per channel for an accelerator at such a large scale. RFSoC data converters that operate in higher-order Nyquist zones demonstrated considerably better RF stability performance than required by C\(^3\). Therefore, we determined that the RFSoC-based platform is the most suitable LLRF solution for C\(^3\), and in this paper a new version of the RFSoC-based NG-LLRF prototype chassis has been introduced. The prototype chassis will be integrated with accelerators not only in the C-band but also in the L-band, S-band, or even X-band with up and down conversion modules for functional and performance evaluation purposes, and the integration procedures and test results will be published. 

The NG-LLRF eliminates the analog RF up and down conversion circuits, which can be challenging to be integrated into the system and maintained after operation. Future versions of NG-LLRF are planned to be designed and implemented with RFSoC on a system-on-module (SOM), so the system can be maintained significantly easier when deployed for C\(^3\) at scale. 

In the paper, we also summarized a selection of high-power test results with pulse train and phase reversal modulation schemes, which are selected among the common amplitude and phase modulation schemes used for RF stations in accelerators. The NG-LLRF demonstrated high flexibility in generating different RF pulses. For instance, it generated and measured a phase flip in 10 ns, which is adequate if used for pulse compression. The NG-LLRF also demonstrated high precision in measuring RF signals at different stages of accelerators. It captured RF pulses with different couplers at high resolution, including transients of RF switching off or phase flipping. These are critical features for characterizing the cavity structures, implementing high precision RF feedback control algorithms, and realizing more exotic RF pulse shaping guided by AI to achieve other operation goals of future accelerators.

AI/ML based control will be essential for achieving the unprecedented performance targets that modern physics experiments require. Conventional LLRF systems and beam dynamics instrumentation suffer from inconsistent data architecture and insufficient throughout, creating a significant technical barrier that prevents aggregation of high quality training and verification datasets across the accelerators for large scale AI/ML control model research and development. The integrated data converters of RFSoC are highly configurable, including the RF frequency, data format, and data rate, so the NG-LLRF can be developed as a common platform for RF signals across accelerators. The RFSoC based system also offers high throughput communication interfaces, such as 40 or 100 GbE, which can be used to stream high fidelity RF data from all the parallel ADC channels. Another technical challenge for implementing large AI/ML control models is the lack of edge computational resources for most conventional LLRF systems. However, the massive digital signal processing compute power integrated in the PL of RFSoC could become the base platform for implementing AI/ML control framework, which will transform raw RF measurements into intelligent action at the edge.

\section*{Acknowledgment}

The work of the authors is supported by the U.S. Department of Energy under Contract No. DE-AC02-76SF00515.

\section*{Data Availability Statement}

The data underlying this article will be shared on reasonable request to the corresponding author.

\nocite{*}
\bibliography{bibliography}

\end{document}